\documentclass[12pt]{article}
\usepackage{amsthm,amsfonts}
\usepackage{amsmath,amssymb}
\usepackage{graphicx}
\usepackage{color}				
\newcommand{\be}{\begin{equation}}
\newcommand{\ee}{\end{equation}}
\newcommand{\bea}{\begin{eqnarray}}
\newcommand{\eea}{\end{eqnarray}}

\newcommand{\und}{\qquad\text{and}\qquad}
\renewcommand{\=}{\ =\ }
\newcommand{\sfrac}[2]{{\textstyle\frac{#1}{#2}}}
\newcommand{\cN}{{\cal N}}
\newcommand{\ic}{\mathrm i}
\newcommand{\diff}{\mathrm d}
\newcommand{\re}{\mathrm{Re}}
\newcommand{\im}{\mathrm{Im}}

\newcommand{\R}{\mathbb R}

\makeatletter
\renewcommand{\@maketitle}{
\newpage
\null
\vskip 1em%
\begin{flushright}
ITP--UH--02/11\\
CBPF-NF-008/11
\end{flushright}
\begin{center}%
{\Large \textbf \@title \par}%
\vskip 2.5em
{\large \@author \par}
\vskip 1.5 em
\end{center}%
\par} \makeatother

\title{An $\cN{=}\,8$ superconformal particle in the half-plane}

\author{Olaf Lechtenfeld$^\ast$ \ and Francesco Toppan$^\dagger$}

\date{}

\begin{document}
\setcounter{page}{0}
\maketitle

\thispagestyle{empty}

\begin{center}
${}^\ast${\it 
Institut f\"ur Theoretische Physik, Leibniz Universit\"at Hannover \\
Appelstra{\ss}e 2, 30167 Hannover, Germany} \\
email: lechtenf@itp.uni-hannover.de

\bigskip

${}^{\dagger}${\it 
CBPF, Rua Dr. Xavier Sigaud 150, Urca \\
cep 22290-180, Rio de Janeiro (RJ), Brazil} \\
email: toppan@cbpf.br
\end{center}

\bigskip

\begin{abstract}
By imposing global supersymmetry and scale invariance we construct 
an $\cN{=}\,8$ superconformal mechanical system based on the 
inhomogeneous (2,8,6) linear multiplet. The unique action describes 
a special K\"{a}hler sigma model with a Calogero-type potential and 
Fayet-Iliopoulos terms. The classical dynamics of the two propagating
bosons is restricted to a (warped) half-plane and bounded. 
We numerically inspect typical trajectories of this special particle.

\end{abstract}

\newpage

\section{Introduction and summary}
For classical mechanics (field theory in $0{+}1$ dimensions) there exists 
a rich landscape of $\cN{=}\,8$ supersymmetric models, distinguished by
the number~$b$ of propagating bosonic degrees of freedom and by the nature of
the supersymmetry transformations (linear or nonlinear)
\cite{BeIvKrLe1,BeIvKrLe2,IvLeSu}. 
Restricting to the linear type, the notation $(b,\cN,\cN{-}b)$ counts their
propagating bosonic, fermionic and auxiliary components.
As was already observed in~\cite{IvKrPa,DoPaJuTs}, an important role is played by 
a potential inhomogeneity in the supersymmetry transformation of the fermions.
The parameters appearing there may be viewed as a constant shift of the auxiliary
components and are introduced through the superfield constraints. Together with
Fayet-Iliopoulos terms, they create a bosonic potential, lead to central charges 
and partial supersymmetry breaking.

To accomodate these inhomogeneous terms, we apply the techniques discussed 
in \cite{PaTo} and~\cite{KuRoTo} and produce the most general inhomogeneous linear 
supermultiplets compatible with the ordinary supersymmetry algebra
$\{Q_i,Q_j\}=\delta_{ij} H$ (without central extensions).

Here, we concentrate on the classical mechanics of a (2,8,6)~particle.
The Lagrangian and Hamiltonian of this model has been formulated for a general
prepotential~$F$ in~\cite{BeKrNe} (without inhomogeneity) and in~\cite{BeKrNeSh} 
(with inhomogeneity).
Here, we specialize to the conformal case and investigate the classical dynamics
of the conformal (2,8,6)~particle.

The inhomogeneous (2,8,6) $\cN{=}\,8$ supermultiplet, under the requirement of 
scale-invariance for the action, defines a unique superconformal mechanical 
system. The only free parameters are the the scale-setting Fayet-Iliopoulos coupling 
and the dimensionless shift entering the inhomogeneous supersymmetry transformations.

We review the inhomogeneous supersymmetry transformations for ${\cal N}{\le}\,8$
and rederive the invariant conformal action for the inhomogeneous (2,8,6) multiplet 
including Fayet-Iliopoulos terms, without using superspace technology. 
After eliminating the auxiliary components we arrive at a very specific 
(non-isotropic and indefinite) Weyl factor and bosonic potential in the two-dimensional 
target space. It proves to be legitimate (at least classically) to restrict to
a (positive-definite) half-space, where we present some typical particle trajectories.

The inhomogeneous supersymmetry transformations that we investigate here 
close the ordinary supersymmetry algebra without central extensions. 
This is the case because we work within the Lagrangian framework. 
Central extensions of the supersymmetry algebra can arise,
both in the classical and quantum cases, as a consequence of 
the Hamiltonian formulation and the closure of the Noether-(super)charge algebra 
under the Poisson bracket structure~\cite{IvKrPa}.

It is tempting to push the idea of this paper to even higher-extended supersymmetry.
For example, by coupling two inhomogeneous (2,8,6) multiplets linked by an extra, 9th,
supersymmetry, one should be able to construct an $\cN{=}\,9$ superconformal mechanics 
model with a four-dimensional target. This might be related with the standard reduction of 
$\cN{=}\,4$ super Yang-Mills to an off-shell multiplet of type (9,16,7) in one dimension.

\newpage

\section{Inhomogeneous minimal linear supermultiplets}
Minimal linear supermultiplets of extended supersymmetry in one dimension are usually
formulated with homogeneous transformations for their component fields. However, in
some cases it is possible to extend the supersymmetry transformations by the addition
of an inhomogeneous term. This is admissible at
\begin{itemize}
\addtolength{\itemsep}{-6pt}
\item $\cN{=}\,2$ for the supermultiplet $(0,2,2)$
\item $\cN{=}\,4$ for the supermultiplets $(0,4,4)$ and $(1,4,3)$
\item $\cN{=}\,8$ for the supermultiplets $(0,8,8)$ and $(1,8,7)$ and $(2,8,6)$
\end{itemize}
The remaining $\cN=2,4,8$ supermultiplets do not admit an inhomogeneous extension,
as can be easily verified by investigating the closure of the 
ordinary $\cN$-extended supersymmetry algebra.
\par
Let $x$ and $y$ be physical bosons, $\psi$, $\psi_i$, $\lambda$ and $\lambda_i$ 
denote fermions, and $g$, $g_i$, $f$ and $f_i$ describe auxiliary fields. 
Here, the isospin index $i$ runs over a range depending on the number of supersymmetries.
The presence of an inhomogeneous term requires the following mass dimension for the fields:
\be
[t]=-1 \qquad\longrightarrow\qquad [x]=-1\ ,\quad [\psi]=-\sfrac12\ ,\quad [g]=0\ . 
\ee
In all the above cases, by a suitable R~transformation, the inhomogeneous terms can be 
rotated to point only in a specific iso-direction. We choose the one with the highest
iso-index, i.e.~$i=2,3$ or~$7$, depending on the case. With this choice, let us list
the various supersymmetry transformations~$Q_i$ for the six cases listed above.

{\bf (0,2,2).}\quad
For the inhomogenous $\cN{=}\,2$ $(0,2,2)$ supermultiplet, the two supersymmetry 
transformations, without loss of generality, can be expressed as
($j,k=1,2$, $\epsilon_{12}=1$)
\bea
&
\begin{array}{ll}
Q_1 \psi_j =g_j\ , &
Q_1 g_j={\dot \psi_j}\ ,
\\[4pt]
Q_2 \psi_j = \epsilon_{jk} {\tilde g_k}\ ,\quad &
Q_2 g_j= \epsilon_{jk}  {\dot \psi_k} \ ,
\end{array}&
\eea
where the inhomogeneous extension hides in
\be
\tilde g_k\ :=\ g_k+c_k \qquad\textrm{with}\quad c_k\in\R\ ,
\ee
and we rotate to $c_1=0$, $c_2\equiv c>0$.

{\bf (0,4,4).}\quad
For the $\cN{=}\,4$ $(0,4,4)$ multiplet, we have ($i,j,k=1,2,3$, $\epsilon_{123}=1$)
\bea
&
\begin{array}{llll}
Q_0\psi=g\ , &Q_0 \psi_j =g_j\ , &
Q_0 g= {\dot \psi}\ , & Q_0 g_j={\dot \psi_j}\ ,
\\[4pt]
Q_i\psi = g_i\ ,\ &Q_i \psi_j = -\delta_{ij} {g}+\epsilon_{ijk} {\tilde g_k}\ ,\ &
Q_i g= -{\dot \psi_i}\ ,&\ Q_i g_j= \delta_{ij} {\dot\psi}-\epsilon_{ijk} {\dot\psi_k}\ ,
\end{array}&
\eea
and we may choose
\be
\tilde g_1=g_1\ ,\quad \tilde g_2=g_2\quad\textrm{but}\quad\tilde g_3=g_3+c\ .
\ee

{\bf (1,4,3).}\quad
The $\cN{=}\,4$ $(1,4,3)$ multiplet looks slightly different,
\bea
&
\begin{array}{llll}
Q_0 x=\psi\ , &Q_0\psi={\dot x}\ , &Q_0 \psi_j =g_j\ , & Q_4 g_j={\dot \psi_j}\ ,
\\[4pt]
Q_i x=\psi_i\ ,\quad& Q_i\psi=-g_i\ ,\quad& 
Q_i\psi_j=\delta_{ij}{\dot x}+\epsilon_{ijk} {\tilde g_k}\ ,\quad&
Q_i g_j= -\delta_{ij} {\dot\psi}-\epsilon_{ijk}  {\dot \psi_k}\ ,
\end{array}&
\eea
with the same $\tilde g_k$ as in (0,4,4).

{\bf (0,8,8).}\quad
Without loss of generality, we can generate the $\cN{=}\,8$ multiplets from the
$\cN{=}\,4$ ones by replacing the quaternionic structure constants~$\epsilon_{ijk}$
by the (totally antisymmetric) octonionic structure constants~$c_{ijk}$, with $i,j,k=1,\ldots,7$ and
\be
c_{123}=c_{147}=c_{165}=c_{246}=c_{257}=c_{354}=c_{367}=1\ ,
\ee
together with $c_{ijk}=0$ for all other index combinations. 
Therefore, the case of (0,0,8) yields
\bea
&
\begin{array}{llll}
Q_0\psi=g\ , &Q_0 \psi_j =g_j\ , & Q_0 g= {\dot \psi}\ , & Q_0 g_j={\dot \psi_j}\ ,
\\[4pt]
Q_i\psi = g_i\ ,\quad&Q_i \psi_j = -\delta_{ij} g+ c_{ijk} {\tilde g_k}\ ,\quad&
Q_i g= -{\dot \psi_i}\ ,\quad & Q_i g_j= \delta_{ij} {\dot\psi}-c_{ijk}  {\dot \psi_k}\ ,
\end{array}&
\eea
and we take
\be
\tilde g_k = g_k + \delta_{k,7}\,c\ .
\ee

{\bf (1,8,7).}\quad
In analogy with (1,4,3), we get
\bea
&
\begin{array}{llll}
Q_0 x=\psi\ , &Q_0\psi={\dot x}\ , &Q_0 \psi_j =g_j\ , & Q_0 g_j={\dot \psi_j}\ ,
\\[4pt]
Q_i x=\psi_i\ ,\quad& Q_i\psi = -g_i\ ,\quad&
Q_i \psi_j = \delta_{ij} {\dot x}+\epsilon_{ijk} {\tilde g_k}\ ,\quad&
Q_i g_j= -\delta_{ij} {\dot\psi}-\epsilon_{ijk}  {\dot \psi_k}\ ,
\end{array}&
\eea
and again $\tilde g_k=g_k$ except for $\tilde g_7=g_7+c$ with $c>0$.

{\bf (2,8,6).}\quad
This is the most interesting multiplet. It is convenient to present it in
quaternionic form, by fusing $(1,4,3)\oplus(1,4,3)=(2,8,6)$, with components
labeled by $(x,\psi_{(i)},g_{(i)})$ and $(y,\lambda_{(i)},f_{(i)})$, respectively,
where $i=1,2,3$.
It is convenient to present the supersymmetry transformations in the following table,
{\tiny
\bea\label{table1}
&
\begin{array}{|c|c|c|c|c|c|c|c|c|c|c|c|c|c|c|c|c|}\hline
&x&g_1&g_2&g_3&y&f_1&f_2&f_3&\psi&\psi_1&\psi_2&\psi_3&\lambda&\lambda_1&\lambda_2&
\lambda_3\\ \hline
Q_0&\psi&{\dot \psi_1}&{\dot\psi_2}&{\dot\psi_3}&\lambda&{\dot\lambda_1}&{\dot\lambda_2}&{\dot\lambda_3}&
{\dot x}&g_1&g_2&g_3&{\dot y}&f_1&f_2&f_3\\\hline
Q_1&\psi_1&-{\dot\psi}&-{\dot\psi_3}&{\dot\psi_2}&\lambda_1&
-{\dot\lambda}&{\dot\lambda_3}&-{\dot\lambda_2}&-g_1&{\dot x}&{\tilde g_3}&-{\tilde g_2}&-f_1&{\dot y}&-{\tilde f_3}&{\tilde f_2}\\\hline
Q_2&\psi_2&{\dot \psi_3}&-{\dot \psi}&-{\dot\psi_1}&\lambda_2&-{\dot\lambda_3}&-{\dot\lambda}&{\dot\lambda_1}&
-g_2&-{\tilde g_3}&{\dot x}&{\tilde g_1}&-f_2&{\tilde f_3}&{\dot y}&-{\tilde f_1}\\\hline
Q_3&\psi_3&-{\dot \psi_2}&{\dot\psi_1}&-{\dot\psi}&\lambda_3&{\dot\lambda_2}&-{\dot\lambda_1}&-{\dot\lambda}&
-g_3&{\tilde g_2}&-{\tilde g_1}&{\dot x}&-f_3&-{\tilde f_2}&{\tilde f_1}&{\dot y}\\\hline
Q_4&\lambda&-{\dot\lambda_1}&-{\dot\lambda_2}&-{\dot\lambda_3}&-\psi&{\dot\psi_1}&
{\dot\psi_2}&{\dot\psi_3}&-{\dot y}&f_1&f_2&f_3&{\dot x}&-g_1&-g_2&-g_3\\\hline
Q_5&\lambda_1&{\dot\lambda}&{\dot \lambda_3}&-{\dot\lambda_2}&-\psi_1&-{\dot\psi}&{\dot\psi_3}&-{\dot\psi_2}&-f_1&-{\dot y}&-{\tilde f_3}&{\tilde f_2}&g_1&{\dot x}&-{\tilde g_3}&{\tilde g_2}\\\hline
Q_6&\lambda_2&-{\dot\lambda_3}&{\dot\lambda}&{\dot\lambda_1}&-\psi_2&-{\dot\psi_3}&
-{\dot\psi}&{\dot\psi_1}&-f_2&{\tilde f_3}&-{\dot y}&-{\tilde f_1}&g_2&{\tilde g_3}&{\dot x}&-{\tilde g_1}\\\hline
Q_7&\lambda_3&{\dot\lambda_2}&-{\dot\lambda_1}&{\dot\lambda}&-\psi_3&
{\dot\psi_2}&-{\dot\psi_1}&-{\dot\psi}&-f_3&-{\tilde f_2}&{\tilde f_1}&-{\dot y}&g_3&-{\tilde g_2}&{\tilde g_1}&{\dot x}\\\hline
\end{array}
&\nonumber
\eea
}
Inspection shows that $Q_0,Q_1,Q_2,Q_3$ act within each of the two (1,4,3) submultiplets,
while the additional supersymmetries $Q_4,Q_5,Q_6,Q_7$ mix the two.
Having SO(3)-rotated inside each (1,4,3) submultiplet to
\be
\tilde g_k = g_k + \delta_{k3}\,c \und \tilde f_k = f_k + \delta_{k3}\,c'
\ee
we may employ a further SO(2) rotation, acting on the $\psi_3\lambda_3$ and
$g_3f_3$ planes, to remove the $c'$ contribution and align the inhomogeneity
with one of the two $\cN{=}\,4$ submultiplets.

\section{Invariant action for a (2,8,6) particle}
To investigate the dynamics of superconformal particles on a line, based on the
various inhomogeneous supermultiplets, we shall need to construct invariant actions
for them. For $\cN{\ge}\,4$ and the presence of at least one physical boson, 
there exists a canonical method~\cite{KuRoTo} to generate such actions, by setting
\be \label{N4action}
{\cal S} \= \int\!\diff t\;{\cal L} \= \int\!\diff t\ Q_1Q_2Q_3Q_4\,F(x,y,\ldots)\ ,
\ee
where $F(x,y,\ldots)$ is an unconstrained prepotential. 
In order to obtain conformally invariant mechanics, the action should not contain
any dimensionful coupling parameter, and therefore, due to $[Q_i]=\sfrac12$, we
demand that $[F]=-1$. One can prove that the ensuing scale invariance extends to
full conformal invariance. 

Without the inhomogeneous extension, (\ref{N4action}) yields only a kinetic term with 
some metric. It is the inhomogeneous term which will give rise to a Calogero-type 
potential.  The action may be complemented by the addition of a Fayet-Iliopoulos term
\be
{\cal S}_{\textrm{FI}} \= \int\!\diff t\;\sum_i(q_i g_i + r_i f_i) 
\qquad\textrm{with}\quad [q_i]=[r_i]=1\ ,
\ee
introducing dimensionful couplings compatible with conformal invariance.
These Fayet-Iliopoulos terms produce an oscillatorial damping, via the DFF 
trick of conformal mechanics~\cite{AlFuFu}.

For the (1,4,3) multiplet (only $x$ and $g_i$, no $y$ or $f_i$), the proper choice
for the prepotential is
\be
F(x)\= \sfrac14\,x\ln x \qquad\longrightarrow\qquad
{\cal L}+{\cal L}_{\textrm{FI}}\=
F''(x)\bigl({\dot x}^2+g_i^2\ +\ c\,g_3\bigr)+q_ig_i
\ +\ \textrm{fermions}\ .
\ee
After eliminating the auxiliary components $g_i$ via their equations of motion
and putting the fermions to zero, one gets
\bea
{\cal L}'_{\textrm{bos}}&=&
F''(x)\bigl({\dot x}^2-\sfrac14c^2\bigr)\ -\ \sfrac14q_i^2/F''(x)\ -\ \sfrac12c\,q_3
\nonumber\\[4pt]
&=&\sfrac14\bigl({\dot x}^2-\sfrac14c^2\bigr)/x\ -\ g_i^2x\ -\ \sfrac12c\,q_3
\label{143pot} \\[4pt]
&=&\sfrac12{\dot w}^2-\sfrac18c^2w^{-2}\ -\ \sfrac12g_i^2w^2\ -\ 
\sfrac12c\,q_3\ ,\nonumber
\eea
and we have recovered the standard conformal action 
after the coordinate change $x=\sfrac12w^2$.

Stepping up to $\cN{=}\,8$, we change the iso-labelling to make $Q_0,Q_1,Q_2,Q_3$
manifest,
\be \label{N8action}
{\cal S} \= \int\!\diff t\;{\cal L} \= \int\!\diff t\ Q_0Q_1Q_2Q_3\,F(x,y,\ldots)\ .
\ee
Demanding invariance under the additional four supersymmetries by requiring
\be \label{N8constraint}
Q_l{\cal L} \= \partial_t W_l \qquad\textrm{for}\quad l=4,5,6,7
\ee
imposes severe constraints on~$F$.
In fact, for the (1,8,7) multiplet no action can be invariant under the inhomogeneous
supersymmetry transformations.\footnote{
In the homogeneous case the constraint reads $F^{\prime\prime\prime\prime}(x)=0$, 
which produces ${\cal L}=(ax{+}b)\,\dot x^2+\ldots$.}

However, the situation is much more interesting for the (2,8,6) multiplet.
Here, the constraint~(\ref{N8constraint}) says that, like in the homogeneous 
case~\cite{GoRoTo}, the prepotential~$F(x,y)$ must be harmonic,
\be
\Box F \ \equiv\ F_{xx}+F_{yy}\=0\ .
\ee
The general solution is encoded in a meromorphic function~$H(z)$ via
\be \label{harmonicF}
F(x,y) \= H(z) + \overline{H(z)} \= 2\,\re H(z)\ ,
\ee
where it is convenient to pass to complex coordinates,
\bea
&
\begin{array}{llll}
z=x+\ic y\ ,\quad& \partial_z=\sfrac12(\partial_x-\ic\partial_y)\ ,\quad&
h_i=g_i+\ic f_i\ ,\quad& \chi_{(i)}=\psi_{(i)}+\ic\lambda_{(i)} \\[4pt]
\bar z=x-\ic y\ ,& \partial_{\bar z}=\sfrac12(\partial_x+\ic\partial_y)\ ,&
\bar h_i=g_i-\ic f_i\ ,& \bar\chi_{(i)}=\psi_{(i)}-\ic\lambda_{(i)}\ .
\end{array}&
\eea

Inserting (\ref{harmonicF}) into~(\ref{N8action}), we obtain
\bea
{\cal L}&=& 
2\,\re\,\bigl\{ H_{zz} (\dot{\bar z}\dot{z}\,+\,\bar h_ih_i\,+\,c\,h_3
\,+\,\sfrac12\dot{\bar\chi}\chi-\sfrac12\bar\chi\dot\chi
\,+\,\sfrac12\dot{\bar\chi}_i\chi_i-\sfrac12\bar\chi_i\dot\chi_i) \nonumber\\[4pt]&+&
H_{zzz} (\chi\chi_ih_i\,+\,\sfrac12\epsilon_{ijk}\chi_i\chi_j h_k\,+\,c\,\chi\chi_3)
\ +\ \sfrac16 H_{zzzz} \epsilon_{ijk}\chi\chi_i\chi_j\chi_k \bigr\}\ ,
\eea
where the inhomogeneous extension is clearly visible in the terms
containing the parameter~$c$.
The bosonic metric $g_{z\bar z}=H_{zz}{+}\bar H_{\bar z\bar z}$ is special
K\"ahler of rigid type~\cite{fre}.
Reverting to real notation and introducing the Weyl factors
\be
\Phi \= 2\,\re H_{zz}\=\sfrac12(F_{xx}{-}F_{yy}) \und
\widetilde\Phi\= -2\,\im H_{zz}\=F_{xy}\ ,
\ee
the Lagrangian reads
\bea
{\cal L}&=&\Phi\bigl({\dot x}^2+{\dot y}^2+{g_i}^2+{f_i}^2-\psi{\dot\psi}
-\lambda{\dot\lambda}-\psi_i{\dot\psi_i}-\lambda_i{\dot\lambda_i}\bigr)
\nonumber\\[4pt]&+& 
\Phi_x\bigl(\psi\psi_ig_i-\psi\lambda_if_i-\lambda\psi_if_i-\lambda\lambda_ig_i\,+\,
\epsilon_{ijk}(\sfrac12 g_i\psi_j\psi_k-\sfrac12 g_i\lambda_j\lambda_k-f_i\lambda_j\psi_k)
\bigr) \nonumber\\[4pt]&+&
\Phi_y\bigl(\lambda\psi_ig_i-\lambda\lambda_if_i+\psi\psi_if_i+\psi\lambda_ig_i\,+\,
\epsilon_{ijk}(\sfrac12 f_i\psi_j\psi_k-\sfrac12 f_i\lambda_j\lambda_k+g_i\lambda_j\psi_k)
\bigr) \nonumber\\[4pt]&+&
\sfrac12(\Phi_{xx}{-}\Phi_{yy})\epsilon_{ijk}\bigl(\sfrac16\psi\psi_i\psi_j\psi_k+\sfrac16\lambda\lambda_i\lambda_j\lambda_k-\sfrac12\psi\psi_i\lambda_j\lambda_k-\sfrac12\lambda\lambda_i\psi_j\psi_k\bigr)
\nonumber\\[4pt]&+&
\Phi_{xy}\,\epsilon_{ijk}\bigl(\sfrac16\lambda\psi_i\psi_j\psi_k-\sfrac16\psi\lambda_i\lambda_j\lambda_k+\sfrac12\psi\lambda_i\psi_j\psi_k-\sfrac12\lambda\psi_i\lambda_j\lambda_k)\bigr)
\nonumber\\[4pt]&+&
c\,\bigl( \Phi g_3 +{\widetilde \Phi}f_3 +\Phi_x(\psi\psi_3-\lambda\lambda_3)+\Phi_y(\lambda\psi_3+\psi\lambda_3) \bigr)\ ,
\eea
to which we add the Fayet-Iliopoulos terms
\be
{\cal L}_{\textrm{FI}} \= q_ig_i+r_if_i\ .
\ee

The harmonic prepotential with the correct scaling dimension~$[H]=-1$ is
\footnote{
Multiplying $H$ with a phase corresponds to an irrelevant rotation in the complex plane.}
\be
H(z) \= \sfrac18\,z\ln z \qquad\longleftrightarrow\qquad
F(x,y)\=\sfrac18\,x\ln(x^2{+}y^2)-\sfrac14\,y\arctan\sfrac{y}{x}\ ,
\ee
and the corresponding Weyl factors read
\be
\Phi\=\sfrac14\,\re\frac1z\=\sfrac14\frac{x}{x^2{+}y^2} \und
\widetilde{\Phi}\=-\sfrac14\,\im\frac1z\=\sfrac14\frac{y}{x^2{+}y^2}\ .
\ee
Note that the corresponding metric is an indefinite one, as it must be for any harmonic
Weyl factor.

In the bosonic limit, obtained by setting all fermions equal to zero, we obtain
\be
{\cal L}_{\textrm{bos}}+{\cal L}_{\textrm{FI}}\=
\Phi\,({\dot x}^2+{\dot y}^2+{g_i}^2+{f_i}^2) +c\,(\Phi\,g_3+{\widetilde\Phi}f_3) 
+q_ig_i+r_if_i\ .
\ee
We eliminate the auxiliary fields via their algebraic equations of motion,
\bea
&
\begin{array}{lll}
g_1=-\frac{q_1}{2\Phi}\ ,\quad& g_2=-\frac{q_2}{2\Phi}\ ,\quad& 
g_3=-\frac{q_3{+}c\Phi}{2\Phi}\ \\[4pt]
f_1=-\frac{r_1}{2\Phi}\ ,\quad& f_2=-\frac{r_2}{2\Phi}\ ,\quad&
f_3=-\frac{r_3{+}c{\widetilde\Phi}}{2\Phi}\ ,
\end{array}&
\eea
and arrive at 
\bea
{\cal L}'_{\textrm{bos}}&=&
\Phi\,\bigl({\dot x}^2+{\dot y}^2\bigr)\ -\ \sfrac1{4\Phi}\bigl(
q_1^2+q_2^2+(q_3{+}c\Phi)^2+r_1^2+r_2^2+(r_3{+}c\widetilde{\Phi})^2\bigr)\nonumber\\[4pt]
&=& \frac{x}{x^2{+}y^2}\frac{{\dot x}^2+{\dot y}^2}{4}\ -\ 
\frac{(q_i^2{+}r_i^2)(x^2{+}y^2)}{x}\ -\ c\,\frac{q_3x{+}r_3y}{2x}\ -\ \frac{c^2}{16x}\\[6pt]
&=:& K\ -\ V\ ,\nonumber
\eea
making explicit the effect of both the inhomogeneous supersymmetry transformation ($c$) and
the Fayet-Iliopoulos terms ($q_i,r_i$) on the potential~$V$.

It is tempting to perform the same coordinate change as for the (1,4,3) multiplet,
$x=\sfrac12w^2$, which yields
\be \label{wy}
{\cal L}'_{\textrm{bos}}\=
\sfrac12(1{+}\gamma^2)^{-1}\Bigl({\dot w}^2+\frac{{\dot y}^2}{w^2}\Bigr)\ -\
\sfrac12(1{+}\gamma^2)(q_i^2{+}r_i^2)w^2\ -\ \sfrac12\,c\,(q_3{+}r_3\gamma)\ -\ 
\frac{c^2}{8w^2}\ ,
\ee
where $\gamma=2y/w^2$. This form reveals both the oscillator and Calogero terms,
but also shows the added complexity in two dimensions (mostly hidden in~$\gamma$).
Putting $y\equiv0$ (also $\gamma{=}0$) brings back the (1,4,3) result~(\ref{143pot}).

\section{Trajectories of a (2,8,6) particle}
Without loss of generality, let us drop inessential Fayet-Iliopoulos terms and put
\be
q_1=q_2=r_1=r_2=0 \und q_3=:q\ ,\quad r_3=:r\ ,\quad q{+}\ic r=:s\ .
\ee
In complex coordinates, the kinetic and potential energies then read
\bea
K&=& \Phi\,{\dot z}\dot{\bar z}\=
\sfrac18\frac{z{+}\bar z}{z\bar z}\,{\dot z}\dot{\bar z}\ ,\\[4pt]
V&=& \bigl( (q{+}c\Phi)^2+(r{+}c\widetilde\Phi)^2\bigr)/4\Phi\=
\sfrac18\frac{1}{z{+}\bar z}\bigl(4s\bar z+c\bigr)\bigl(4\bar s z+c\bigr)\ .
\eea 
\begin{figure}[ht]
\centerline{
\lower2ex\hbox{
\includegraphics[width=9cm]{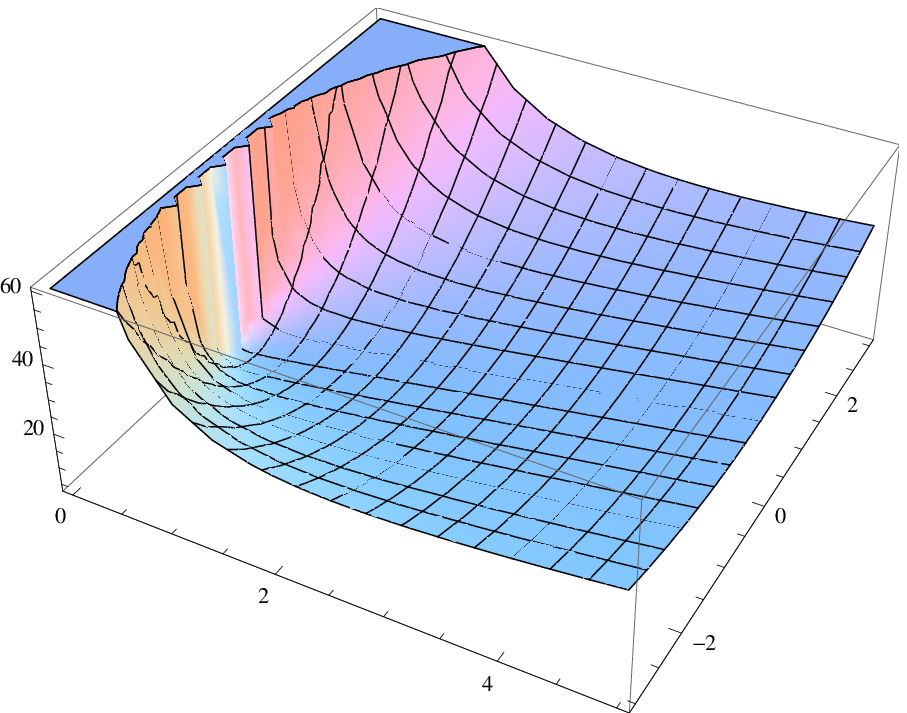}
}
\hfill
\includegraphics[width=5cm]{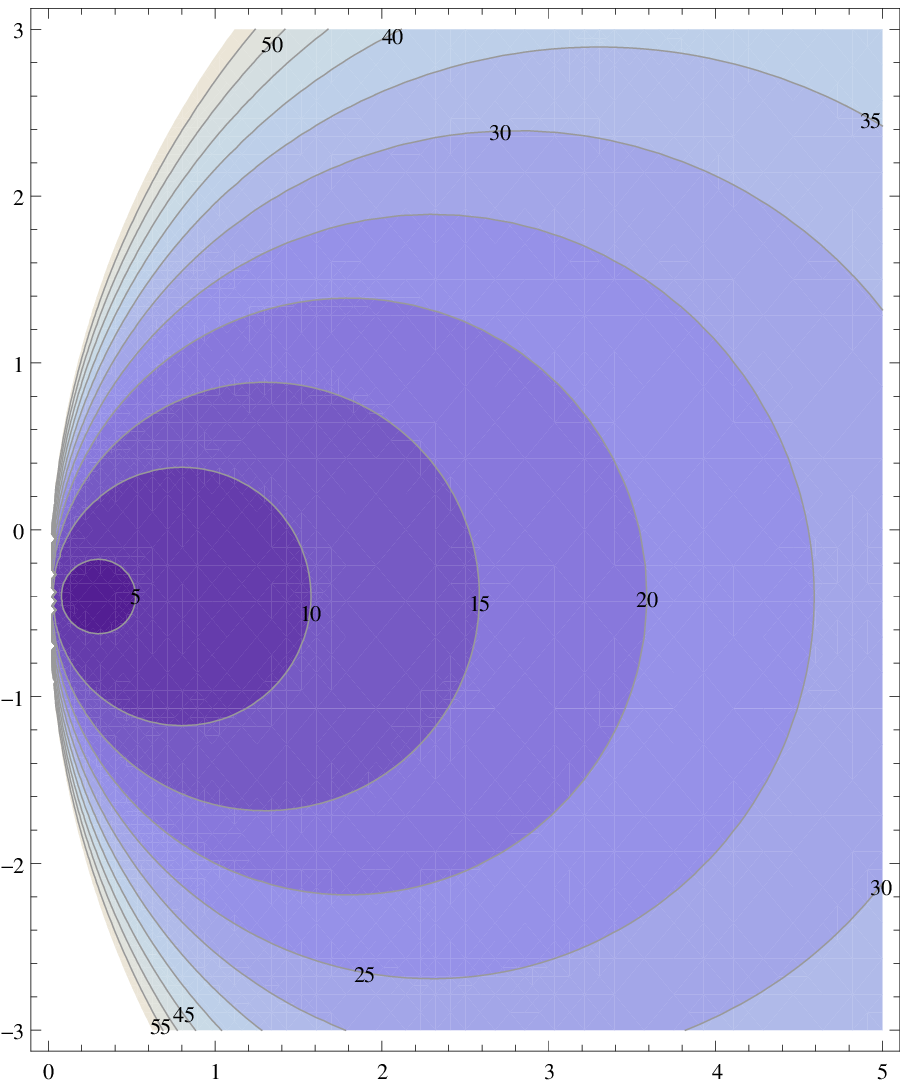}
}
\caption{Potential~$V$ and its level curves for $(c,q,r)=(4,1,2)
\quad\longrightarrow\quad z_{\textrm{min}}=\sfrac15(1{-}2\ic)$. }
\label{fig:1}
\end{figure}
The level curves of this potential are circles of center and radius
\be
z_0(V)=\frac{2V-c\,s}{4(q^2{+}r^2)} \und
r(V)=\frac{\sqrt{V(V{-}c\,q)}}{2(q^2{+}r^2)}\ ,
\ee
respectively, and its only minimum $V_{\textrm{min}}=cq$ is located at
\be
z_{\textrm{min}}=z_0(cq)=\frac{c\,\bar s}{4(q^2{+}r^2)}\ .
\ee
The parameter~$r$ governs the asymmetry under $y\to-y$.
The reflection $x\to-x$ flips the sign of $V{-}\sfrac12cq_3$.
Due to the factor of $z{+}\bar z=2x$, both the Weyl factor and the potential are
strictly positive on the right half-space $x{>}0$ and strictly negative for $x{<}0$.
Therefore, the (2,8,6) particle is a reasonable dynamical system only if its trajectories
do not cross the $x{=}0$ dividing line.
Seen from the right half-space, the potential barrier for $x{\to}0$ has a hole at $y{=}0$
if $c{=}0$, but the Weyl factor explodes precisely there.
For large coordinate values, the potential grows linearly with~$x$ and 
quadratically with~$y$, so the $x{>}0$ trajectories remain bounded.

The equation of motion takes the form
\bea
0&=&\Phi^3\ddot z\ +\ \Phi^2\Phi_z {\dot z}^2\ -\
\sfrac14\Phi_{\bar z}\bigl(q^2+(r+2\ic cH_{zz})^2\bigr) \nonumber\\[4pt]
&\propto &(z{+}\bar z)^3 z\bar z\,\ddot z\ -\ (z{+}\bar z)^2 \bar z^2{\dot z}^2\ +\
z^2\bar z^2\bigl( (4qz)^2+(4rz{+}\ic c)^2\bigr)\ ,
\eea
which in real coordinates reads
\bea
0&=& \ddot x-\frac{1}{2x}\frac{x^2{-}y^2}{x^2{+}y^2}({\dot x}^2{-}{\dot y}^2)
-\frac{2y}{x^2{+}y^2}\,\dot x\,\dot y
+\frac{x^2{+}y^2}{x^3}\bigl(2(q^2{+}r^2)(x^2{-}y^2)-cr\,y-\sfrac18 c^2\bigr)
\ ,\nonumber\\[4pt]
0&=& \ddot y+\frac{y}{x^2{+}y^2}({\dot x}^2{-}{\dot y}^2)
-\frac{1}{x}\frac{x^2{-}y^2}{x^2{+}y^2}\,\dot x\,\dot y
+\frac{x^2{+}y^2}{x^3}\bigl(4(q^2{+}r^2)\,x\,y+cr\,x\bigr)\ .
\eea
The only constant of motion of this system is the energy $E=T+V$, so the generic particle
motion is not integrable. Figure~2 shows the trajectory for the $(c,q,r)$-value
chosen in Figure~1 and a couple of initial conditions.
\begin{figure}[ht]
\centerline{
\includegraphics[width=7.7cm]{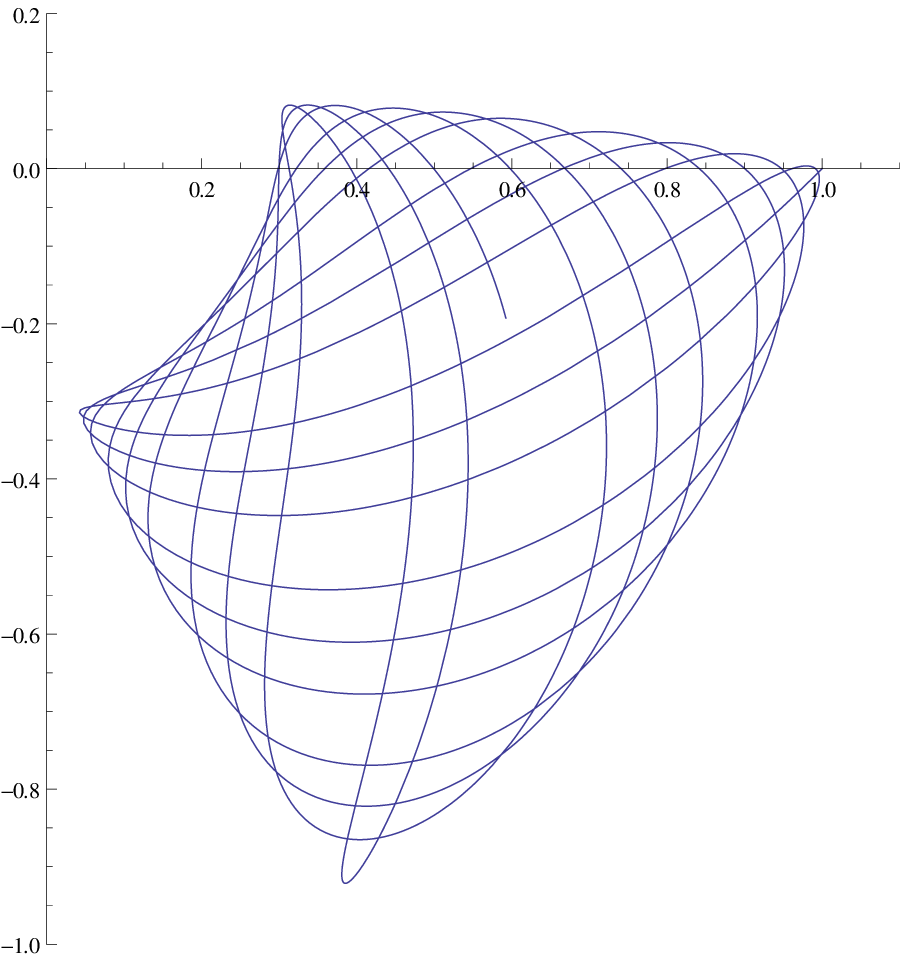}
\hfill
\includegraphics[width=6.3cm]{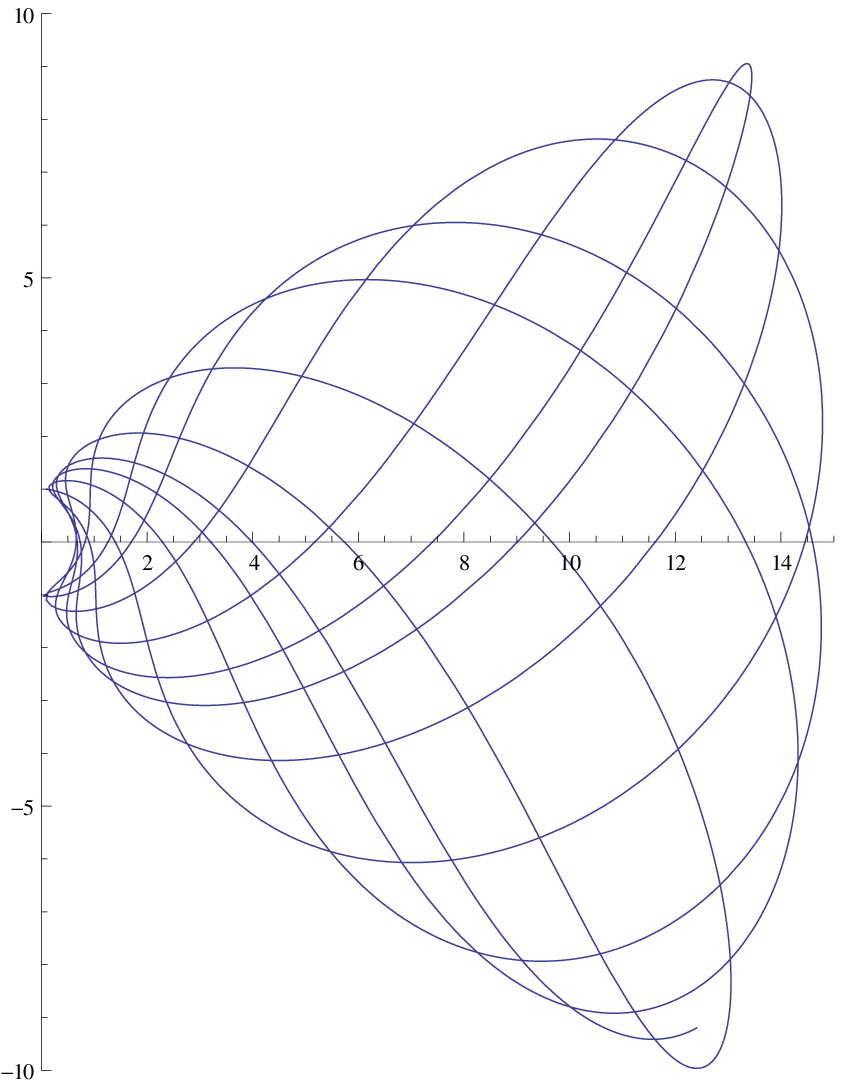}
}
\caption{Trajectories for $(c,q,r)=(4,1,2)$ with initial conditions
$(z,\dot z)(0)=(1,0)$ (left) and $(z,\dot z)(0)=(\sfrac{1}{10}{+}\ic,0)$ (right). }
\label{fig:2}
\end{figure}
One sees that the curve does not fill out the region~$V(x,y)\le E$, 
on effect of the position-dependent effective mass $M=2\Phi(x,y)$.
It is also clear that the $x{=}0$ barrier is impenetrable.
Therefore, it makes sense to substitute $w=\sqrt{2x}$ and introduce the dynamics 
in the $wy$-plane according to~(\ref{wy}). The trajectories of Figure~2 get
somewhat distorted in these variables, but their qualitative behavior is unchanged.

\bigskip

\noindent
{\bf Acknowledgements}

\noindent
O.L. thanks CBPF for warm hospitality. This work was partially supported by 
CNPq and by DFG grant Le-838/9-2.


\end{document}